\documentclass[sigconf,screen]{acmart}
\usepackage{todonotes}
\usepackage{booktabs}
\usepackage{listings}
\usepackage{natbib} 
\usepackage{comment}
\usepackage{graphicx}
\usepackage{xcolor}
\usepackage[many]{tcolorbox} 
\usepackage{tablefootnote}

\definecolor{gray50}{gray}{.5}
\definecolor{gray40}{gray}{.6}
\definecolor{gray30}{gray}{.7}
\definecolor{gray20}{gray}{.8}
\definecolor{gray10}{gray}{.9}
\definecolor{gray05}{gray}{.95}

\tcbset{
    sharp corners,
    colback = white,
    before skip = 0.2cm,    
    after skip = 0.5cm      
}                           
\newtcolorbox{boxC}{
    colback = sub, 
    boxrule = 0pt  
}

\definecolor{main}{HTML}{CFCFCF}    
\definecolor{sub}{HTML}{CFCFCF}     

\begin{document}



\title{Initial Insights on MLOps: Perception and Adoption by Practitioners} 

 

\author{Sergio Moreschini}
\affiliation{
  \institution{University of Oulu}
  \city{Oulu}
  \country{Finland}
}
\email{sergio.moreschini@oulu.fi}

\additionalaffiliation{
     \institution{Tampere University, Tampere, Finland, sergio.moreschini@tuni.fi} 
     }
    
\author{David Hästbacka}
\affiliation{
  \institution{Tampere University}
  \city{Tampere}
  \country{Finland}
}

\email{david.hastbacka@tuni.fi}

\author{Andrea Janes}
\affiliation{
  \institution{Free University of Bozen-Bolzano}
  \city{Bozen-Bolzano}
  \country{Italy}
}
\email{andrea.janes@unibz.it}

\author{Valentina Lenarduzzi}
\affiliation{
  \institution{University of Oulu}
  \city{Oulu}
  \country{Finland}
}

\email{valentina.lenarduzzi@oulu.fi}

\author{Davide Taibi}
\affiliation{
  \institution{University of Oulu}
  \city{Oulu}
  \country{Finland}
}

\email{davide.taibi@oulu.fi}







\begin{abstract}
The accelerated adoption of AI-based software demands precise development guidelines to guarantee reliability, scalability, and ethical compliance. MLOps (Machine Learning and Operations) guidelines have emerged as the principal reference in this field, paving the way for the development of high-level automated tools and applications. Despite the introduction of MLOps guidelines, there is still a degree of skepticism surrounding their implementation, with a gradual adoption rate across many companies. In certain instances, a lack of awareness about MLOps has resulted in organizations adopting similar approaches unintentionally, frequently without a comprehensive understanding of the associated best practices and principles. 
The objective of this study is to gain insight into the actual adoption of MLOps (or comparable, e.g. AIOps) guidelines in different business contexts. To this end, we surveyed practitioners representing a range of business environments to understand how MLOps is adopted and perceived in their companies. The results of this survey also shed light on other pertinent aspects related to the advantages and challenges of these guidelines, the learning curve associated with them, and the future trends that can be derived from this information.
This study aims to provide deeper insight into MLOps and its impact on the next phase of innovation in machine learning. By doing so, we aim to lay the foundation for more efficient, reliable, and creative AI applications in the future.
\end{abstract}


\maketitle

\section{Introduction}

The past decade has witnessed an AI revolution, with significant shifts in how industries approach data utilization. Two domains stand out as key catalysts for this transformation:  generative AI (GenAI) and machine learning operations (MLOps)~\cite{intro1}. While both domains offer unprecedented opportunities for revolutionizing industries, their roles and applications differ significantly. 

GenAI represents a category of AI that is capable of generating new content~\cite{intro3}. Some applications, such as AI assistants, are gaining interest for their ability to assist developers by automating routine coding tasks, increasing productivity, and reducing technical debt~\cite{intro4}. GenAI is positioned as a valuable tool for the software development lifecycle, particularly due to its ability to provide real-time assistance and generate high-quality code snippets. These features underscore its potential to drive transformative change across a range of technology and industry sectors. By offering rapid assistance and streamlined coding procedures, GenAI becomes an invaluable \textit{partner} in software development, particularly when aligned with MLOps guidelines.

MLOps provides a robust framework for managing the ML lifecycle, addressing key challenges such as scalability, reliability, and collaboration between data scientists and operations teams. Understanding these aspects enables more efficient and effective implementation of ML projects, ultimately accelerating innovation and providing a competitive edge to organizations, paving the way for more efficient, reliable, and creative AI applications in the future. 

MLOps combines machine learning and DevOps to streamline the machine learning lifecycle, from initial development through to deployment and maintenance. The ultimate goal of MLOps is to create scalable and reliable ML systems that can be integrated seamlessly into existing workflows. By automating repetitive tasks and ensuring robust monitoring and management of models in production, MLOps promises to enhance the efficiency and effectiveness of ML projects~\cite{intro2}.

Despite the growing interest and investment in these technologies, there remains a significant gap in understanding how MLOps is perceived and implemented in the real world, particularly from an industrial perspective. There is currently a high level of skepticism about MLOps guidelines and the importance of using them when developing ML-based software. Therefore, the objective of this work is to deepen the understanding of MLOps and its role in driving the next wave of machine learning innovation.
To achieve this, we aim to address the aforementioned gap in knowledge and demonstrate the importance of such guidelines by capturing the experiences and insights of practitioners from diverse organizational contexts. This will highlight both the practical challenges and benefits of adopting MLOps. 

We designed and conducted a qualitative survey among three MLOps experts investigating the perspectives of developers and leaders in large enterprises and SMEs, uncovering the implications of MLOps adoption. It addresses key issues such as tool costs, data drift, and the daunting learning curve associated with MLOps, as well as the potential for advanced tools powered by Generative AI to address these challenges. 

We can summarize our main findings as follows:
\begin{itemize}

	\item Some developers are unknowingly following MLOps guidelines.
	\item The high cost of tools poses a barrier to experimenting with MLOps.
	\item The daunting learning curve for MLOps practices is primarily due to mindset challenges.
	\item Generative AI is on the verge of becoming a valuable support tool for developers.
\end{itemize}

The rest of the work is organized as follows: Section~\ref{sec:Design} presents the design of the empirical study. It includes multiple subsections, beginning with the introduction of the goal and research questions (RQs). We will then explain how the survey has been designed and how it is structured. Finally, we will describe how the study was performed and analyzed, and how it can be replicated.
Section~\ref{sec:Perspective} reports the perspective of the interviewees on each of the different aspect questioned.
Section~\ref{sec:Discussion} presents the results and discuss them based on each RQ.
Section~\ref{sec:RelatedWorks} provides other relevant works within the same topic.
Section \ref{sec:Conclusion} draws the relevant conclusions.  
\section{Empirical Study}
\label{sec:Design}
In this Section, we describe our empirical study designed as a survey according to the Wohlin et al. guideline~\cite{Wohlin2012}.

\subsection{Goal and Research Questions}

The goal is to raise awareness about the current use of MLOps in organizations. By comparing the different approaches used by different stakeholders, we aim to understand not only the current state of MLOps adoption but also the associated learning curve and stakeholders' vision of future applications and opportunities.

Therefore, we derived the following Research Questions (RQ). 

\begin{boxC}
\textbf{RQ$_1$.} To what extent do developers understand and embrace the concept of MLOps?
\end{boxC}

\textbf{RQ$_1$} is designed to assess developers' familiarity with MLOps principles, their implementation practices, and the extent to which they have integrated these practices into their workflow. In particular, it seeks to determine whether developers are actively adopting MLOps practices or if these practices are being incorporated indirectly as a result of applying machine learning (ML) within DevOps frameworks.

\begin{boxC}
\textbf{RQ$_2$.} What improvements could enhance MLOps by leveraging existing benefits and solving existing issues for pipelines and tools?
\end{boxC}

In \textbf{RQ$_2$} we aim to identify the key improvements needed to enhance MLOps practices by leveraging current benefits and addressing existing issues in pipeline development and tools. Additionally, it seeks to uncover the factors that are impeding the adoption and awareness of MLOps among practitioners.

\begin{boxC}
\textbf{RQ$_3$.} What is the perception of the MLOps learning curve in the industry?
\end{boxC}

\textbf{RQ$_3$ }aims to gain insight into how practitioners perceive the difficulty of adopting MLOps and to assess how hard is to learn the correct practicess. By exploring practitioners’ experiences and perspectives, we can identify the specific challenges that contribute to the perceived difficulty and how these challenges impact the adoption process.

\begin{boxC}
\textbf{RQ$_4$.} What are the anticipated future trends in MLOps?
\end{boxC}

Lastly, in \textbf{RQ$_4$}, we aim to elicit insights from participants on the tools and practices that are likely to gain prominence and the directions in which the field is expected to evolve. We are interested in understanding their views on the development of advanced tools with a comprehensive understanding of the entire MLOps pipeline.


\subsection{Study Design}
To answer our research questions, we designed and conducted a qualitative survey consisting of the five aspects described below. 

\begin{enumerate}
    \item \textit{MLOps adoption (RQ$_1$)}. The first part of the survey is related to the current adoption of MLOps. As a starting point, we aim to understand the user's experience in the context of MLOps and its position on the classical MLOps model in literature (such as the one from Google \cite{Google}). In addition, it is important to understand what are the main reasons and applications for the adoption of MLOps within the engineering team, as well as the number of people that make up such a team.

\item \textit{MLOps architecture (RQ$_1$)}. The second part focuses on tools and phases. Specifically, we aim to understand which phases constitute the pipeline of the practitioner, compared to the well-known DevOps infinite loop representation. We also aim to determine the importance given to each phase in terms of priority and automation. The tools are subject to the same types of questions, not only in terms of related phases but also regarding timing, satisfaction (scale 1-5), and potential (scale 1-5). Questions about known but unused tools and related feedback loops conclude this section. 

\item \textit{MLOps issues and benefits (RQ$_2$)}. The third section is a direct consequence of the second, as at first issues and benefits of different tools are compiled following questions from the previous section. The other questions composing this section are focused on the whole MLOps pipeline and include benefits, issues, and possible improvements.

\item \textit{Comparison of MLOps pipelines (RQ$_3$)}. The purpose of this section is to compare the current state of MLOps adoption by practitioners with our vision of an ideal MLOps pipeline. After presenting our ideal pipeline, we asked for suggestions and ideas on how to improve it, as well as opinions on the division of roles between software developers and ML developers. We also asked about any difficulties in learning how to apply the proposed pipeline.

\item \textit{Future trends (RQ$_4$)}. The last section aims to understand the practitioners' vision of possible future trends concerning applications or tools. 
\end{enumerate}

In the following text, we describe the participants selection and the questions we formulated to them.

\subsection{Questionnaire}
The five sections outlined incorporate a combination of \textbf{twentyfour questions} designed as open-ended and Likert scale items with a range of 1-5 (where 1 indicates a complete lack of likeability and 5 indicates complete likeability).

\subsubsection{Participants Selection}
Three participants were selected for this study based on several criteria, including their diverse professional backgrounds. This was done to ensure a comprehensive perspective on the subject matter. 
The purpose of this selection of participants is to illustrate the various challenges and methodologies encountered in diverse organizational contexts and technical domains. MLOps and AI researchers and developers working in large enterprises provide insights into the complexities of integrating systems within vast, data-rich environments. In contrast, software developers at small to medium-sized enterprises (SMEs) offer a view of the agile and innovative techniques required for developing ML/AI algorithms within a constrained and rapidly evolving setting.

The contrast between these backgrounds is crucial to understanding the broader landscape of software engineering practices and how they can best be used in a range of different environments. By examining these different viewpoints, this study will be able to provide valuable insights and highlight potential best practices that can be used in a wide variety of organizations and for different types of technical work.

The \textbf{first participant} is an experienced researcher and developer employed by a large enterprise. This participant has a wealth of experience from working within a larger organizational structure. This participant’s expertise lies specifically in recommender systems. This is an area that involves sophisticated algorithms and user data interactions to personalize and enhance user experiences.

The \textbf{second participant} is a software developer working in an SME. Given the resource constraints and dynamic nature of smaller organizations, a more versatile and adaptive approach to software development is often required in this setting. The participant's specialization is in algorithm development for ML and AI applications, focusing on the creation and optimization of algorithms that underpin intelligent systems.

The \textbf{third participant} is the research and development department vice president of an SME. This role encompasses a broad range of responsibilities, including involvement in multiple projects as a project manager and hands-on development contributor due to the SME’s collaborative and dynamic nature. The participant’s extensive experience in overseeing and practically contributing to the development of various projects provides a unique perspective on balancing strategic oversight with practical implementation in a resource-constrained environment.

\subsubsection{MLOps adoption (RQ$_1$)} We formulated \textbf{6 questions} to assess the participant’s familiarity with the MLOps concept and to determine whether the participant consciously and intentionally implements MLOps practices or if these practices have been adopted inadvertently, without explicit recognition of the MLOps framework.
It is crucial to understand this distinction, as it helps differentiate between the deliberate, informed adoption of MLOps methodologies and incidental usage that may occur due to overlapping practices in machine learning and software engineering. By grasping this differentiation, we can gain valuable insights into the level of awareness and intentionality behind MLOps adoption, which is essential for assessing the effectiveness and integration of these practices within different organizational contexts.

The subsequent questions were selected based on the participant's initial response. If the participant demonstrated familiarity with and intentional use of MLOps practices, the following questions incorporated specific MLOps terminology. Conversely, if the participant does not have knowledge on the  terminology related to MLOps but used such practices without referring them as MLOps, the interviewer continued the interview by using the "DevOps for ML" terminology. This document will focus on the terminology used in the first case.


This approach guarantees that the language used in the survey is consistent with the participant’s level of expertise and familiarity, thus enhancing the clarity and relevance of the questions. It permits a more tailored investigation of the participant’s practices and experiences, whether explicitly framed within the MLOps paradigm or understood through the more general context of DevOps applied to ML. This differentiation is essential for accurately capturing the participant’s perspectives and the specific practices they employ in their work.
In light of the participant’s demonstrated knowledge of MLOps, the subsequent question was tailored to further explore their understanding of MLOps standards and practices. Conversely, for those less familiar with MLOps, the focus was on understanding if any machine learning (ML) approach was used. If this question yielded a negative response, the interview would conclude, resulting in an unsuccessful outcome.

\subsubsection{MLOps architecture (RQ$_1$)} we formulated \textbf{8 questions} to ascertain knowledge and evaluate perceptions. This method aims to first ascertain participants’ understanding of specific aspects of MLOps architecture and then assess their perceptions of the importance of these aspects.

For instance, the initial question in this section focused on identifying the various phases within the MLOps lifecycle. Following this, the  participants were asked to evaluate the significance of each identified phase. This two-step approach ensures a comprehensive exploration of both the participant’s knowledge and their evaluative perspective on key components of MLOps architecture.

By structuring the questions in this manner, the survey can effectively capture detailed insights into how participants perceive and prioritize different phases of MLOps, providing valuable data for understanding the adoption and implementation of MLOps practices across different organizational contexts.

In particular (question 3.3), we asked the participant to describe the tools used in each phase of their MLOps processes, as well as to evaluate these tools in terms of their satisfaction and perceived potential. By gathering detailed information about the specific tools employed and the context of their use within the MLOps lifecycle, this question aimed to obtain valuable insights into the tools being utilized in the MLOps process. Moreover, Question 3.3 served as a direct link to the subsequent section. For each tool mentioned, participants were also asked to detail the benefits and issues associated with it in Question 4.4. This connection between the questions ensures a comprehensive understanding of the tools’ practical applications and effectiveness, as well as the challenges encountered during their implementation.

By correlating the descriptions of tools with their evaluated benefits and issues, the survey aims to provide a comprehensive view of the current tool landscape in MLOps. This approach allows for a more in-depth analysis of how different tools contribute to the MLOps workflow and potential areas for improvement in tool development and integration.

\subsubsection{MLOps issues and benefits (RQ$_2$)} We formulated \textbf{4 questions} to gain insights into potential areas for improvement within the MLOps processes and tools. We intentionally linked the final question directly to the previous section. Consequently, some participants may have already addressed it.

Specifically, (question 4.3) we requested that participants consider how their current MLOps practices could be improved. This methodology enabled the survey to obtain a comprehensive range of perspectives on existing challenges and potential areas for enhancement. By encouraging participants to identify and discuss both the advantages and the issues they encountered, the survey aimed to provide a balanced and in-depth understanding of the current MLOps landscape. The feedback gathered in this section is both actionable and relevant, and it highlights specific areas where MLOps practices can be refined to better meet the needs of practitioners. 

\subsubsection{Comparison of MLOps pipelines (RQ$_3$)} we designed  \textbf{4 questions}, each related to a figure illustrating an MLOps infinite loop pipeline. The \textit{Comparison of MLOps Pipelines} section aimed to directly compare the interviewee’s conception of MLOps with the pipeline model proposed in our previous work~\cite{MLOpsFigure}. 

The questions prompted participants to assess and compare their understanding and application of MLOps with the proposed pipeline model. By doing so, the survey aimed to identify similarities and differences in MLOps practices, offering valuable insights into the diversity of approaches within the field.

The open-ended nature of these questions enabled participants to provide detailed feedback on how their MLOps practices align with or diverge from the proposed model. This approach facilitated a deeper understanding of the practical application of MLOps and highlighted areas where our proposed pipeline might be improved or adapted to better fit real-world scenarios.

\subsubsection{Future trends (RQ$_4$)} In The final section,  we asked the participants to answer to \textbf{2 questions}, each intended to prompt informed speculation about the anticipated evolution of MLOps tools and areas of focus. During the survey, we encouraged participants to think forward and share their visions of how MLOps will evolve. The insights gained from these discussions offer invaluable foresight into the potential trajectory of the industry.

The initial question sought to identify which tools, or categories of tools, the participants believe will become more prominent or widely adopted in the future. This question aimed to gain insights into emerging technology and innovations that participants see as having significant potential to influence the future of MLOps practices. The second question was designed to identify the specific areas or applications within MLOps that participants anticipate will receive greater focus and development. By understanding which aspects of MLOps participants believe will become priorities, this question provides valuable insight into future trends and areas of growth within the field.

\subsection{Survey Execution}
We collected the information through open-ended questions, running the survey as a face-to-face interview. Two authors collected the answers separately and then checked possible inconsistencies in the report. The disagreements were discussed and clarified with the other authors. We informed the participants, according to the GDPR\footnote{https://gdpr-info.eu}, about their rights and that they could abandon the study anytime. Moreover, all information provided by each participant has been treated as confidential, without disclosing any sensible data.

\subsection{Data Analysis}
The qualitative data analysis has been conducted individually by each author.
Moreover, pairwise inter-rater reliability was measured across the three sets of decisions to get a fair/good agreement on the first iteration of this process. 
Based on the disagreements, we clarified possible discrepancies and different classifications. A second iteration resulted in 100\% agreement.

\subsection{Replicability and Verifiability}
To allow the verifiability and the replicability of our study, we publish all the raw data including the transcript of each interview in the online appendix\footnote{\url{dx.doi.org/10.6084/m9.figshare.26408197}\label{Package}}. For privacy reasons, we anonymized the names of the companies.

\section{Results}
\label{sec:Perspective}
In this section, we report the obtained results grouped by the three experts interviewed. The answers will be summarized for each aspect of the questionnaire

\begin{table*}[t]
\centering
\begin{tabular}{l|l|l|l|l|l}
\hline
\textbf{Tool}           & \textbf{Phase}  & \textbf{Satisfaction} & \textbf{Potential} & \textbf{Benefit}               & \textbf{Issue}                            \\ \hline
AWS SageM.\footnotemark    & 1-3     & 2            & 3         & Fully-managed ML Lifecycle        & Requires Expertise                \\
EMR\footnotemark           & 1-3      & 2            & 3         & Low Costs                 & Not exactly ML/Requires Expertise \\ 
Airflow\footnotemark        & 2         & 4            & 4         & Intuitive UI/Community          & Not exactly ML/Requires Expertise                   \\ 
Databricks\footnotemark     & 2         & 5            & 5         & Easy/Comprehensive Solution            & High Costs                             \\ 
MLflow\footnotemark          & 2         & 5            & 5         & Track/Reproducibility & Setup Cost                       \\ 
ElasticSearch\footnotemark   & 2-3       & 4            & 5         & Fast Learning Pace         & Setup Cost + Production Cost     \\ 
Presto\footnotemark        & 2-3       & 4            & 5         & Fast Learning Pace            & Setup Cost + Production Cost                   \\ 
DataDog\footnotemark        & 4         & 4            & 5         & Available Dashboards             & Setup cost/Requires Expertise/Self-Managed \\ 
AWS CloudWatch\footnotemark  & 4        & 4            & 5         & Fast Learning Pace                  & Non-intuitive UI                               \\ 
HoneyComb\footnotemark      & 4         & 4            & 5         & Intuitive UI                    & Specific Syntax                  \\ \hline
\end{tabular}
\caption{Tools Summary First Interviewee}
\label{tab:Tools1}
\end{table*}

\subsection{First Interview}

The first interview took place on February 20th, 2024, and lasted about 45 minutes. The interviewee, an ML engineer, focuses on full-stack research and development at his company (Large Enterprise). His main field of expertise is Recommender Systems (RSs) with more than 5 years of experience.

\subsubsection{MLOps adoption} The first interviewee claimed extensive experience with MLOps frameworks, especially in the areas of deployment and debugging. The user-developed frameworks have been used in the area of RSs (specifically for TV channels and hotel recommendations)  to improve the user experience. These frameworks do not adhere to a specific standard but are tailored to the requirements of each project. His participation in various projects across different fields has resulted in fluctuations in team sizes, depending on the specific requirements of each application. However, the size of such teams has always been fewer than five people.

\subsubsection{MLOps architecture} When asked about the different phases composing the MLOps architecture, the interviewee identified 4 main phases of equal importance: 
    \begin{enumerate}
        \item Prototyping
        \item Offline Training
        \item Online Model
        \item Monitoring
    \end{enumerate}
    Such phases rely on different tools which have been summarized in Table~\ref{tab:Tools1}. From the table it can be seen that no tool is capable of covering all the different phases and a combination of these is necessary to achieve full MLOps. The most important of these tools are those used in the monitoring stage (as they are linked to the various ongoing notebooks to create the feedback loop), MLFlow, and Databricks. However, even though Databricks has the potential to be the most satisfying, it has been replaced with a combination of Airflow and a monitoring system for cost savings. 
    The interviewee also stated that Phases 2 and 3 are most important to automate first and that the main reason for retraining can be identified in Data Drift.

\footnotetext[3]{\url{https://aws.amazon.com/sagemaker/}}
\footnotetext[4]{\url{https://aws.amazon.com/emr/}}
\footnotetext[5]{\url{https://airflow.apache.org}}
\footnotetext[6]{\url{https://www.databricks.com}}
\footnotetext[7]{\url{https://mlflow.org}}
\footnotetext[8]{\url{https://www.elastic.co/elasticsearch}}
\footnotetext[9]{\url{https://prestodb.io}}
\footnotetext[10]{\url{https://www.datadoghq.com}}  
\footnotetext[11]{\url{https://aws.amazon.com/cloudwatch/}}  
\footnotetext[12]{\url{https://www.honeycomb.io}}  

\subsubsection{MLOps issues and benefits} When starting the discussion related to the main issues in MLOps the interviewee stated: 

\vspace{2mm}
\begin{quote}
        \textit{``MLOps is used to prevent issues''.}
    \end{quote} 

\vspace{2mm}
It is however its nature of continuous improvement and the continuous addition of features and functionalities that create the most issues especially when creating new APIs. Another valuable point is also the difficulty of performing multiple ML iterations without increasing the technical debt. Despite these issues, however, the main benefits of MLOps are evident especially when analyzing the high quality of the ML model and the provided User Experience. More importantly, these systems are far from being perfect and they necessitate high effort, especially for the Offline phase. Such a phase requires high synchronization between the different teams within a company working on different projects. The main issues and benefits of tools are also reported in Table~\ref{tab:Tools1}.

\subsubsection{Comparison of MLOps pipelines} When presented with our vision the interviewee noticed similarities in every phase except for the deployment phase. He suggested specifying the different levels kind of possible deployment, namely \textit{Local, Developer Environment, Online and Smoke Test}. Moreover, when specifically asked about the binomial between SE and ML engineer he stated that it is an \textit{Utopian} representation, as such binomial would mostly create "translation errors" due to the missing knowledge that one of the figures might have on the project. Regarding the adoption of such models, he believes that the most challenging part is related to the mindset of the Non-ML practitioners which need to see the benefits of applying MLOps, while on the practical side, the main challenge is the creation of the whole infrastructure that would allow a full automation.
    
\subsubsection{Future trends} When discussing future trends, the interviewee foresees an increased effort in creating LLMs for code writing support, also known as Code Copilot. The interviewee expressed excitement for the future development of a similar application applied to unit tests. For what concerns the areas that will gain the most attention he identified Healthcare and Explainable AI.

\begin{table*}[h]
\centering
\begin{tabular}{l|l|l|l|l|l}
\hline
\textbf{Tool}           & \textbf{Phase}  & \textbf{Satisfaction} & \textbf{Potential} & \textbf{Benefit}               & \textbf{Issue}                            \\ \hline
Elixir (P. Language)\footnotemark    & Development    &    Not Provided         & Not Provided         &      None    & Low Maturity               \\
Neovim (IDE)\footnotemark            & Development      & Not Provided           & Not Provided         & Low Costs                 & Not exactly ML/Requires Expertise \\ 
Huggingface\footnotemark        & Training         & Not Provided             & Not Provided         & UI/Community          & Not exactly ML/Requires Expertise                   \\ 
Amazon SageMaker   & Training         & Not Provided           & Not Provided          & Resources Release            &          None                    \\ 
Amazon EC2 \footnotemark    & Training         & Not Provided            & Not Provided          &         None    &      Idle Costs                      \\ 
Amazon S3 \footnotemark        & Deployment         & Not Provided            & Not Provided        & Allows Separation & None                      \\  \hline
\end{tabular}
\caption{Tools Summary Second Interviewee}
\label{tab:Tools2}
\end{table*}

\subsection{Second Interview}
The second interview took place on March 14th, 2024, and lasted about 1 hour. The interviewee has a completely different background than the first one, he is a software developer who is dedicated to the development of algorithms for the application of ML and AI within the company he works for (Small/Medium Enterprise).

\subsubsection{MLOps adoption} The second interviewee, when asked about MLOps, claimed to have heard of such practices.
    To ease the interview process, the interviewer provided an MLOps definition~\cite{MLOpsPrinciples}. The interviewee was unaware of any specific MLOps standards and had never used them in a project. On the contrary, their models are too heavy to be part of the final product, so their deployed code points to a file repository. This allows them to deploy the product code and the associated model separately, thereby reducing the dependency of one on the other. Specifically, when a new model is created, the developed application only needs to be pointed to the new model address. Once the model is loaded, the startup process of the application begins. 

    In this case, the division between the model (ML side) and the code (SE side) is more evident, reducing also the level of automation. Interviewees identified this division as the primary reason for MLOps, further specifying that most model development, testing, and validation are done offline without any level of automation, and automation is only initiated in the monitoring phase after deployment. Explicitly: 
    
    \vspace{2mm}
    \begin{quote}
        \textit{``Implementing an MLOps architecture'' – even if the model development is not in a pipeline – means to be able to deploy code and model separately.}
    \end{quote}
    \vspace{2mm}
    
    The developed solutions have been used in speech-to-text applications and the team is composed of less than 10 people (including administrative and management).

\footnotetext[13]{\url{https://elixir-lang.org}}
\footnotetext[14]{\url{https://neovim.io}}
\footnotetext[15]{\url{https://huggingface.co}}
\footnotetext[16]{\url{https://aws.amazon.com/pm/ec2/}}
\footnotetext[17]{\url{https://aws.amazon.com/s3/}}
    
\subsubsection{MLOps architecture} When asked about the different phases composing the architecture the interviewee referred to the previously provided definition of MLOps~\cite{MLOpsPrinciples}, specifically to its maturity level. The identified maturity level was the first as most of the ML-related tasks are performed manually but the upcoming goal is to move to level 2 by obtaining human-corrected text to perform continuous training and deployment. With the first level of maturity, the interviewee identified as the most critical phase of the testing in production. Explicitly: 
    
        \vspace{2mm}
        \begin{quote}
        \textit{``Understand if the latest deployed model is not causing regression''.}
    \end{quote}
        \vspace{2mm}
        
    Also, in this case, the architectural approach relies on the utilization of several tools, as summarised in Table~\ref{tab:Tools2}. Nevertheless, the satisfaction and potential of these tools have not been evaluated by this participant.  In this level of maturity, to reach the next level, the interviewee identified the deployment phase as the most important supported by automated testing and continuous training.

    When discussing the feedback mechanisms and the consequent reasons for performing retraining, the interviewee identified two feedback events, the first defined as ``catastrophic events" caused by the client identifying an issue with the application and the model quality. Specifically, the second reason for the model degradation is identified in data drift.

\subsubsection{MLOps issues and benefits} The main issue identified by the second interviewee is the cost related to performing MLOps. Specifically, those related to tool usage (Amazon EC2 has been abandoned due to idle costs) and those related to retraining. In this interview the importance of performing correct retraining has been highlighted: i.e. retraining optimization to avoid useless costs but also to improve the model when necessary. The main benefit of the approach used has been identified in the quality of the product provided which consequently increases customer satisfaction.

In this context, the interviewer asked what could be improved, and the interviewee expressed a high level of satisfaction with the proposed solution, which currently does not need to be updated. However, the future perspective of this solution is that if there is a new and more accentuated data drift, it may be necessary to restructure the approach, including automation, to increase the level of maturity.

\begin{table*}[t]
\centering
\begin{tabular}{l|l|l|l|l|l}
\hline
\textbf{Tool}           & \textbf{Phase}  & \textbf{Satisfaction} & \textbf{Potential} & \textbf{Benefit}               & \textbf{Issue}                            \\ \hline
Jenkins\footnotemark   & Deployment    &    3        &  4        &      Fast Learning Pace    & None               \\
SonarQube\footnotemark           & Development      & 2           & 2         & Fast Learning Pace                 & Limited Features \\ 
Azure\footnotemark       & Ops         & 4           & 5        & Relatively Fast Pace Learning          & Costly/Daunting Learning Curve                   \\ 
Grafana\footnotemark     & Monitoring         & 3           & 5          & Fast Learning Pace            &          Tool Unstable                   \\ 
Nifi\footnotemark     & Ops         & 5            & 5          &         Fast Learning Pace    &      None                      \\ 
SparkML\footnotemark        & Analysis         & 3 & 4        & Fast Learning Pace & None                      \\ 
Hadoop\footnotemark         & Monitoring         & 3 & 3        & Moderate Pace Learning & Hard to Configure                      \\  \hline
\end{tabular}
\caption{Tools Summary Third Interviewee}
\label{tab:Tools3}
\end{table*}

\subsubsection{Comparison of MLOps pipelines} Again, we presented our vision, and the first reaction was triggered by the concept of parallelizing. For the interviewee, our vision required two people working almost synchronously, while in their approach it was important to reduce the dependency between the model and the code. Instead of having two figures working at the same \textit{speed}, they needed two processes that could be updated independently. He suggested explicitly considering the importance of traceability within the proposed infinite loop to increase explainability. He was very pragmatic about adopting the model, stating that it is unlikely to get adopted because they only implement what is necessary.

When asked about the comparison between SE engineers and ML engineers, he pointed out the absence of another important figure in the picture: the hardware engineer. The interviewee regards the hardware engineer as an expert in GPU and training infrastructure, without whom the other two cannot create a realistic plan.

\subsubsection{Future trends} When discussing future trends, the second interviewee believes that monitoring tools in the future will be able to monitor the entire MLOps pipeline, including continuous training before deployment. This would enable optimization in the coding phase based on feedback received at each stage, both before and after deployment.
In terms of areas that will receive more attention in the future, the respondent believes that hardware accelerators will become more important based on current development trends.

\subsection{Third Interview}
The third interview was conducted on July 17th, 2024, and lasted approximately 20 minutes. The interviewee's background differs from that of the previous two interviewees. He is the vice-director of the research and development department of the company where he is employed (SME and Consultancy). 

\footnotetext[18]{\url{https://www.jenkins.io}}
\footnotetext[19]{\url{https://www.sonarsource.com/products/sonarqube/}}
\footnotetext[20]{\url{https://azure.microsoft.com}}
\footnotetext[21]{\url{https://grafana.com}}
\footnotetext[22]{\url{https://nifi.apache.org}}
\footnotetext[23]{\url{https://spark.apache.org/docs/latest/ml-guide.html}}
\footnotetext[24]{\url{https://hadoop.apache.org}}

\subsubsection{MLOps adoption} In response to questions about his experience with MLOps, the third interviewee asserted that he has been utilizing MLOps practices for approximately two years in three distinct project contexts. His approach to strategy and adoption was consistently tailored to the specific projects he worked on, necessitating a correspondingly adapted pipeline to align with project-specific needs.
Consequently, the primary rationale for implementing MLOps was to facilitate the continuous analysis and monitoring of data generated within the aforementioned projects. These projects were primarily related to administrative matters concerning cities and regions within the domain of smart cities. The primary objective was the development of a recommendation system for users (i.e., citizens) that would facilitate the discovery of events within their respective cities or regions.
The project development team consisted of ten individuals. Of these, five were software developers. Two were focused on IT operations, and three were assigned to specific tasks. One was responsible for the monitoring phase, one was in charge of the planning phase, and one was engaged in data analysis.

\subsubsection{MLOps architecture} In response to a question regarding the various phases that comprise the architectural framework, the interviewee identified the DevOps infinite loop as a fundamental element. He asserted that every distinctly developed pipeline was based on this loop and encompassed all the requisite phases.
The aforementioned phases were all identified as equal, except for the planning phase, which was accorded a lesser importance due to the straightforward nature of the project requirements. The group of phases composing the Ops was considered the most important to automate first.
Following expectations, the MLOps pipelines developed in this case also relied on the utilization of several tools, as summarized in Table~\ref{tab:Tools3}. As reported, the tools exhibited an overall average performance and met the anticipated standards. Notable exceptions included NiFi, which demonstrated the highest degree of satisfaction, while Jenkins, SparkML, and particularly Grafana, demonstrated potential for improvement. Furthermore, Kafka Stream was originally included in a project but was subsequently excluded due to data distribution issues, with MapReduce being replaced by SparkML. The rationale behind this latter decision was the limited number of phases that MapReduce was able to offer.
In terms of the feedback mechanism, this was entirely contingent on the report of the data analysis. Following this, the report was then provided to both developers and clients. Subsequently, clients would then proceed to provide a second feedback response to the developers, which would then be based on both the original report and the subsequent client feedback.
In their methodology, the data assumes a pivotal role, and given its inherent variability, two potential triggers—data drift and the one-month expiration—have been identified as the basis for retraining.
    
\subsubsection{MLOps issues and benefits} The discussion on the subject of issues related to MLOps proceeded to address the social implications and general acceptance of this concept, rather than focusing on the practical aspects involved in the utilization and advancement of pipelines. 
In particular, the interviewee highlighted the absence of a consensus surrounding the concept of MLOps as a primary factor deterring investment from stakeholders and companies regarding MLOps. This ultimately leads to a shortage of individuals adequately trained in the concept, which further distances the concept from such realities. To mitigate these issues, the interviewee proposes a focus on the interaction between human and MLOps pipelines, to facilitate the explainability of the pipeline. Nevertheless, the advantages of timeliness and an uncomplicated work schedule are significant when MLOps are utilized. 
    
When the focus is on the tools (as detailed in Table~\ref{tab:Tools3}), the most significant benefit and desired quality of specific tools is the ease of use and configuration. While various types of issues have been identified for different tools, the one that presented the greatest challenges was Azure, which was perceived as both costly and having a "steep learning curve." During the interview, the concept of a "steep learning curve" was introduced as a metaphor that likely reflects a common perception that navigating a steep hill is more challenging than traversing a long, gradual incline~\cite{curve}. To not cause any potential confusion, it will then be reported as daunting.
   
\subsubsection{Comparison of MLOps pipelines} In this instance, when our vision was presented to the interviewee, it was observed that the proposed MLOps pipeline exhibited certain similarities with their developing pattern. In response to the question regarding potential improvements, he once more emphasized the necessity for procedures that facilitate explainability and identified the ML debug phase as the optimal context for such improvements.
    
The following question pertained to the complexity of transitioning to an infinite loop. In alignment with the preceding discussion, he highlighted resource constraints as a key challenge. He proposed a classification of these resource constraints based on their human and economic dimensions. From a human perspective, there is a notable lack of knowledge, while from an economic standpoint, the cost of tools such as Azure reduces the likelihood of implementing these approaches through experimentation. Consequently, the initial investment is likely to be high before any return is seen.

The final question of this section addressed the potential for collaboration between software engineers and those with expertise in machine learning (ML). The interviewee not only endorsed the dualism but also proposed the addition of a third role. This third role, which would be filled by an individual with a background in theoretical software development, would ensure adherence to best practices and reduce technical debt.
    
\subsubsection{Future trends} The discussion on prospective trends centered on the significance of LLMs and Generative AI. The interviewee suggested that the application of this new technology will facilitate the coding phase. He anticipates that the tools will enable the automation of mechanical development processes. Furthermore, the introduction of LLM will enhance the debugging process, leading to more transparent code and a reduction in the time required for debugging.

\section{Discussion}
\label{sec:Discussion}
\begin{table*}[h]
\centering
\begin{tabular}{l|l|l|l}
\hline
    & Interviewee 1                  & Interviewee 2                               & Interviewee 3                              \\ \hline
RQ$_1$ & Experienced user - RSs         & Indirect user - Speech-to-text applications & Experienced user - RSs                     \\
RQ$_2$ & Cost reduction (tools)         & Cost reduction (tools)                      & Improved Human-MLOps pipelines interaction \\
RQ$_3$ & Daunting learning curve - mindset & Almost impossible - mindset                 & Daunting learning curve - mindset             \\
RQ$_4$ & GenAI for coding support & Optimization through improved monitoring tools & GenAI for coding support \\ \hline
\end{tabular}
\caption{Interviewees Summary for different RQs}
\label{tab:Interviews}
\end{table*}
In this Section, we summarize and discuss the obtained results answering to our RQs.

\subsection{MLOps concept understandability (RQ$_1$)} 
The interviews detailed in the previous section provide a comprehensive overview of perspectives on MLOps, highlighting distinct viewpoints influenced by organizational size and maturity in adopting MLOps practices. One participant from a large organization and one from an SME explicitly implemented MLOps to enhance automation within their workflows. In contrast, the remaining participants from an SME employed MLOps practices indirectly and at a lower maturity level. Despite these differences, all projects were facilitated through the strategic use of diverse tools tailored to specific goals. 
One notable insight from the interviews was regarding the existence of an "offline" phase, where models are trained and tested before deployment without automation assistance. This practice highlights the critical importance of model reliability, which cannot always be represented by a single objective metric, like accuracy. For example, automated pipelines that use tools like MLflow~\cite{mlflow} for continuous deployment must consider multiple evaluation criteria that go beyond the use of basic measures.

\subsection{MLOps improvement (RQ$_2$)} 
The interviews revealed a consensus on the dual nature of MLOps, presenting both significant benefits and notable challenges. All participants reported that adopting MLOps practices enhanced product quality and increased client satisfaction. However, the cost of tools was identified as a potential obstacle, prompting some to switch to more cost-effective alternatives. Furthermore, data drift was identified as a pervasive issue, necessitating frequent model retraining to maintain performance and relevance.

\subsection{MLOps perception (RQ$_3$)} 
A common theme across all interviews was the learning curve associated with adopting MLOps, primarily influenced by the need for a mindset shift among practitioners and stakeholders. This shift often required a catalyst, such as the promise of increased benefits or meeting specific organizational requirements. In many cases, the anticipated higher return on investment served as this trigger, though it often took considerable time to realize.

\subsection{MLOps future trends (RQ$_4$)} 
Considering future developments, participants speculated on the potential for more advanced tools with a comprehensive understanding of the entire MLOps pipeline. These future tools aim to provide enhanced support to developers, either by improving the coding and testing phases or by offering holistic pipeline feedback mechanisms. Participants envisioned these advancements driven by the evolution of Generative AI, which could also propel the development of Explainable AI, thereby enhancing accessibility and interpretability within the MLOps field.

A summary of the individual perspective on the different RQs is reported in Table~\ref{tab:Interviews}. The table included in the study highlights that at least two out of the three interviewees agreed on each key point of the research questions (RQs). This consensus among the participants serves to validate the relevance and accuracy of the RQs, demonstrating that the questions were well-formulated to capture essential insights into the benefits, challenges, and future directions of MLOps practices.

\section{Related Works}
\label{sec:RelatedWorks}
Several studies have attempted to define guidelines for MLOps from various standpoints. In one of our previous work~\cite{MLOpsFigure}, we outlined our view of MLOps guidelines as a logical progression of DevOps. This involved transforming the traditional infinite loop into a new model where software engineers and ML engineers work closely together.

Among the most widely recognized and universally accepted are the guidelines proposed by Google. In their Google Cloud Architecture Center blog~\cite{Google}, researchers have outlined different levels of automation to classify maturity in MLOps. This framework provides a structured approach to scaling MLOps, ranging from manual processes to fully automated CI/CD pipelines.

Additionally, Sculley et al.\cite{HiddenSculley} from Google presents a vision of ML systems through the lens of MLOps, emphasizing the importance of accounting for technical debt in the software development process~. This work illustrates how failing to address technical debt can result in significant and prolonged maintenance challenges, ultimately leading to reduced system performance. It is essential to consider this perspective to gain insight into the long-term implications of MLOps adoption and the necessity for robust, scalable solutions.


Other works have been focusing on the importance of the taxonomy related to MLOps and other similar terms. Steidl et al.~\cite{Monika} compared terms such as DevOps and CI/CD for AI, MLOps, and Continuous Delivery for Machine Learning (CD4ML). Moreover, they investigated the factors that could trigger retraining or pipeline execution, as well as the potential challenges associated with MLOps development. However, Ianniello et al.~\cite{taxonomy} conducted a review of the existing scientific literature and proposed a taxonomy for clustering research papers on MLOps. Among the findings of the work, is that the application of DevOps principles to ML and the use of MLOps in industrial environments remain relatively under-researched in academic literature. 

Matsui et al.~\cite{ResponsibleAI} identified 5 steps to guide researchers and practitioners toward understanding and adopting MLOps. The first and most important step in the process is to encourage a mindset that is open to learning and change. They highlight the importance of cultural aspects for the successful implementation of MLOps, noting that without a shift in mindset and organizational culture, technical advancements alone are insufficient for achieving effective MLOps adoption.
Other works have specifically focused on challenges.

Diaz-de-Arcaya et al.~\cite{ChallengesSurvey} provides a comprehensive survey of the challenges and opportunities in both MLOps and AIOps (AI for IT Operations). This study presents a roadmap for future developments in these fields, emphasizing the necessity for more robust integration and collaboration between MLOps and AIOps practices. They conducted a comprehensive literature review, automating the article retrieval process to identify key areas for improvement and potential advancements. Their work offers valuable insights, with one key finding being that MLOps is particularly beneficial in challenging environments, such as industry, while AIOps excels in challenging circumstances, such as 5G and 6G. 

Faubel et al.~\cite{MLOpsIndustry} also identify challenges as a key topic in the context of MLOps for Industry 4.0. The authors distinguish three primary categories of tasks that present challenges, namely data, models, and support activities. Notably, support activities are situated between the former two categories, highlighting their unique characteristics. Consequently, the four support activities identified in the article—namely, infrastructure, tools, versioning, and automation—are regarded as the most challenging. 

On a completely different perspective described in the review performed by Haertel et al.~\cite{DataProject}, MLOps is seen as a solution to the development and production of ML models specifically for Data Science (DS) projects. The systematic literature review of 52 papers on MLOps for DS revealed that, although the academic field is still in its early stages, approximately half of the work describes practical use cases. Furthermore, while MLOps literature primarily addresses technical aspects, it often overlooks organizational considerations. Achieving success in this field depends on the effective management of both analytical and technical expertise.

Another valuable SLR is the one presented in John et al~\cite{MLOpsMaturity}, which in this case is supported also by a grey literature review. The reviews have led to the creation of a framework and maturity model, which have subsequently been validated in three different company cases. The cases demonstrate that the companies interviewed were still in the first two phases of the presented maturity model. This indicates that while two of the companies were automating data collection, a third company was also incorporating deployment into its automation pipeline. However, none of those were implementing any kind of automation for monitoring. 

Finally in John et al~\cite{Tradeoffs} the authors conducted an extensive multi-case analysis involving nine professionals from seven different companies to uncover the key trade-offs that organizations face when adopting MLOps. This study systematically categorized these trade-offs using the BAPO model, which stands for Business, Architecture, Process, and Organization. Each of these categories represents a critical dimension of MLOps adoption, reflecting the multifaceted nature of integrating MLOps practices into existing workflows. The results of this multi-case analysis not only highlighted these trade-offs but also led to the identification of possible mitigations. These mitigations are strategies that organizations can employ to balance the competing demands of the BAPO dimensions. 
\section{Conclusion}
\label{sec:Conclusion}
This research study presents a survey on the existing use of MLOps in various organizational settings. 

The interviews yielded insights into the varying perspectives on MLOps adoption, influenced by organizational size and maturity. The interviewees offered varying perspectives as users, providing insights that were both consistent and divergent. Participants from a large enterprise and an SME utilized MLOps explicitly for automation, while another SME participant employed it indirectly at a lower maturity level. All leveraged diverse tools tailored to specific goals, emphasizing the significance of tool diversity.

While continuous training and deployment are not yet fully developed ideas, MLOps practices have been shown to improve product quality and client satisfaction. However, high tool costs and data drift have posed significant challenges, with the latter specifically necessitating frequent model retraining.

The learning curve presented a significant challenge, often requiring a shift in mindset that was prompted by the promise of increased benefits or specific organizational needs. The introduction of advanced tools with a comprehensive understanding of the MLOps pipeline, driven by Generative and Explainable AI, is anticipated to enhance developer support.

In summary, the study identifies the necessity for enhanced integration and support tools within MLOps, addresses key challenges such as the cost of tools and data drift, and emphasizes the importance of a cultural shift to facilitate smoother MLOps adoption. Addressing these issues can lead to greater awareness and adoption of MLOps within organizations.

\section*{Acknowledgment}

This work is funded by IndustryX (Business Finland), 6GSoft projects (Business Finland), and Mufano project (Business Finland).


\bibliographystyle{ACM-Reference-Format}
\bibliography{Manuscript}


\begin{thebibliography}{19}


\ifx \showCODEN    \undefined \def \showCODEN     #1{\unskip}     \fi
\ifx \showDOI      \undefined \def \showDOI       #1{#1}\fi
\ifx \showISBNx    \undefined \def \showISBNx     #1{\unskip}     \fi
\ifx \showISBNxiii \undefined \def \showISBNxiii  #1{\unskip}     \fi
\ifx \showISSN     \undefined \def \showISSN      #1{\unskip}     \fi
\ifx \showLCCN     \undefined \def \showLCCN      #1{\unskip}     \fi
\ifx \shownote     \undefined \def \shownote      #1{#1}          \fi
\ifx \showarticletitle \undefined \def \showarticletitle #1{#1}   \fi
\ifx \showURL      \undefined \def \showURL       {\relax}        \fi
\providecommand\bibfield[2]{#2}
\providecommand\bibinfo[2]{#2}
\providecommand\natexlab[1]{#1}
\providecommand\showeprint[2][]{arXiv:#2}

\bibitem[Andreev(2022)]%
        {curve}
\bibfield{author}{\bibinfo{person}{Ivan Andreev}.} \bibinfo{year}{2022}\natexlab{}.
\newblock \bibinfo{title}{What is steep learning curve?}
\newblock \bibinfo{howpublished}{\url{https://www.valamis.com/hub/steep-learning-curve}}.
\newblock


\bibitem[Center(2023)]%
        {Google}
\bibfield{author}{\bibinfo{person}{Google Cloud~Architecture Center}.} \bibinfo{year}{2023}\natexlab{}.
\newblock \bibinfo{title}{MLOps: Continuous delivery and automation pipelines in machine learning}.
\newblock \bibinfo{howpublished}{\url{https://cloud.google.com/architecture/mlops-continuous-delivery-and-automation-pipelines-in-machine-learning}}.
\newblock


\bibitem[Diaz-de Arcaya et~al\mbox{.}(2023)]%
        {ChallengesSurvey}
\bibfield{author}{\bibinfo{person}{Josu Diaz-de Arcaya}, \bibinfo{person}{Ana~I. Torre-Bastida}, \bibinfo{person}{Gorka Z\'{a}rate}, \bibinfo{person}{Ra\'{u}l Mi\~{n}\'{o}n}, {and} \bibinfo{person}{Aitor Almeida}.} \bibinfo{year}{2023}\natexlab{}.
\newblock \showarticletitle{A Joint Study of the Challenges, Opportunities, and Roadmap of MLOps and AIOps: A Systematic Survey}.
\newblock \bibinfo{journal}{\emph{ACM Comput. Surv.}} \bibinfo{volume}{56}, \bibinfo{number}{4} (\bibinfo{year}{2023}).
\newblock
\showISSN{0360-0300}


\bibitem[Esposito et~al\mbox{.}(2024)]%
        {intro4}
\bibfield{author}{\bibinfo{person}{Matteo Esposito}, \bibinfo{person}{Andrea Janes}, \bibinfo{person}{Davide Taibi}, {and} \bibinfo{person}{Valentina Lenarduzzi}.} \bibinfo{year}{2024}\natexlab{}.
\newblock \showarticletitle{Generative AI in Evidence-Based Software Engineering: A White Paper}.
\newblock \bibinfo{journal}{\emph{arXiv preprint arXiv:2407.17440}} (\bibinfo{year}{2024}).
\newblock


\bibitem[Esposito and Palagiano(2024)]%
        {intro3}
\bibfield{author}{\bibinfo{person}{Matteo Esposito} {and} \bibinfo{person}{Francesco Palagiano}.} \bibinfo{year}{2024}\natexlab{}.
\newblock \showarticletitle{Leveraging Large Language Models for Preliminary Security Risk Analysis: A Mission-Critical Case Study}. In \bibinfo{booktitle}{\emph{Proceedings of the 28th International Conference on Evaluation and Assessment in Software Engineering}}. ACM, \bibinfo{pages}{442--445}.
\newblock


\bibitem[Faubel et~al\mbox{.}(2023)]%
        {MLOpsIndustry}
\bibfield{author}{\bibinfo{person}{Leonhard Faubel}, \bibinfo{person}{Klaus Schmid}, {and} \bibinfo{person}{Holger Eichelberger}.} \bibinfo{year}{2023}\natexlab{}.
\newblock \showarticletitle{MLOps Challenges in Industry 4.0}.
\newblock \bibinfo{journal}{\emph{SN Computer Science}} \bibinfo{volume}{4}, \bibinfo{number}{6} (\bibinfo{year}{2023}), \bibinfo{pages}{828}.
\newblock


\bibitem[Godwin and Melvin(2024)]%
        {intro2}
\bibfield{author}{\bibinfo{person}{Ryan~C Godwin} {and} \bibinfo{person}{Ryan~L Melvin}.} \bibinfo{year}{2024}\natexlab{}.
\newblock \showarticletitle{Toward efficient data science: A comprehensive MLOps template for collaborative code development and automation}.
\newblock \bibinfo{journal}{\emph{SoftwareX}}  \bibinfo{volume}{26} (\bibinfo{year}{2024}), \bibinfo{pages}{101723}.
\newblock


\bibitem[Haertel et~al\mbox{.}(2023)]%
        {DataProject}
\bibfield{author}{\bibinfo{person}{Christian Haertel}, \bibinfo{person}{Daniel Staegemann}, \bibinfo{person}{Christian Daase}, \bibinfo{person}{Matthias Pohl}, \bibinfo{person}{Abdulrahman Nahhas}, {and} \bibinfo{person}{Klaus Turowski}.} \bibinfo{year}{2023}\natexlab{}.
\newblock \showarticletitle{MLOps in Data Science Projects: A Review}. In \bibinfo{booktitle}{\emph{2023 IEEE International Conference on Big Data (BigData)}}. \bibinfo{pages}{2396--2404}.
\newblock
\urldef\tempurl%
\url{https://doi.org/10.1109/BigData59044.2023.10386139}
\showDOI{\tempurl}


\bibitem[John et~al\mbox{.}(2021)]%
        {MLOpsMaturity}
\bibfield{author}{\bibinfo{person}{Meenu~Mary John}, \bibinfo{person}{Helena~Holmström Olsson}, {and} \bibinfo{person}{Jan Bosch}.} \bibinfo{year}{2021}\natexlab{}.
\newblock \showarticletitle{Towards MLOps: A Framework and Maturity Model}. In \bibinfo{booktitle}{\emph{2021 47th Euromicro Conference on Software Engineering and Advanced Applications (SEAA)}}. \bibinfo{pages}{1--8}.
\newblock


\bibitem[John et~al\mbox{.}(2023)]%
        {Tradeoffs}
\bibfield{author}{\bibinfo{person}{Meenu~Mary John}, \bibinfo{person}{Helena~Holmström Olsson}, \bibinfo{person}{Jan Bosch}, {and} \bibinfo{person}{Daniel Gillblad}.} \bibinfo{year}{2023}\natexlab{}.
\newblock \showarticletitle{Exploring Trade-Offs in MLOps Adoption}. In \bibinfo{booktitle}{\emph{2023 30th Asia-Pacific Software Engineering Conference (APSEC)}}. \bibinfo{pages}{369--375}.
\newblock


\bibitem[L’Esteve(2023)]%
        {intro1}
\bibfield{author}{\bibinfo{person}{Ron~C L’Esteve}.} \bibinfo{year}{2023}\natexlab{}.
\newblock \showarticletitle{Impacts of modern AI and ML trends}.
\newblock In \bibinfo{booktitle}{\emph{The Cloud Leader’s Handbook: Strategically Innovate, Transform, and Scale Organizations}}. \bibinfo{publisher}{Springer}, \bibinfo{pages}{135--155}.
\newblock


\bibitem[Matsui and Goya(2023)]%
        {ResponsibleAI}
\bibfield{author}{\bibinfo{person}{Beatriz M.~A. Matsui} {and} \bibinfo{person}{Denise~H. Goya}.} \bibinfo{year}{2023}\natexlab{}.
\newblock \showarticletitle{MLOps: a guide to its adoption in the context of responsible AI}. In \bibinfo{booktitle}{\emph{Proceedings of the 1st Workshop on Software Engineering for Responsible AI}} \emph{(\bibinfo{series}{SE4RAI '22})}. \bibinfo{pages}{45–49}.
\newblock


\bibitem[Moreschini et~al\mbox{.}(2022)]%
        {MLOpsFigure}
\bibfield{author}{\bibinfo{person}{Sergio Moreschini}, \bibinfo{person}{Francesco Lomio}, \bibinfo{person}{David Hästbacka}, {and} \bibinfo{person}{Davide Taibi}.} \bibinfo{year}{2022}\natexlab{}.
\newblock \showarticletitle{MLOps for evolvable AI intensive software systems}. In \bibinfo{booktitle}{\emph{2022 IEEE International Conference on Software Analysis, Evolution and Reengineering (SANER)}}. \bibinfo{pages}{1293--1294}.
\newblock
\urldef\tempurl%
\url{https://doi.org/10.1109/SANER53432.2022.00155}
\showDOI{\tempurl}


\bibitem[Projects(2023)]%
        {mlflow}
\bibfield{author}{\bibinfo{person}{LF Projects}.} \bibinfo{year}{2023}\natexlab{}.
\newblock \bibinfo{title}{MLflow - A platform for the machine learning lifecycle}.
\newblock \bibinfo{howpublished}{\url{https://www.mlflow.org}}.
\newblock


\bibitem[Sculley et~al\mbox{.}(2015)]%
        {HiddenSculley}
\bibfield{author}{\bibinfo{person}{David Sculley}, \bibinfo{person}{Gary Holt}, \bibinfo{person}{Daniel Golovin}, \bibinfo{person}{Eugene Davydov}, \bibinfo{person}{Todd Phillips}, \bibinfo{person}{Dietmar Ebner}, \bibinfo{person}{Vinay Chaudhary}, \bibinfo{person}{Michael Young}, \bibinfo{person}{Jean-Francois Crespo}, {and} \bibinfo{person}{Dan Dennison}.} \bibinfo{year}{2015}\natexlab{}.
\newblock \showarticletitle{Hidden technical debt in machine learning systems}.
\newblock \bibinfo{journal}{\emph{Advances in neural information processing systems}}  \bibinfo{volume}{28} (\bibinfo{year}{2015}).
\newblock


\bibitem[Steidl et~al\mbox{.}(2023)]%
        {Monika}
\bibfield{author}{\bibinfo{person}{Monika Steidl}, \bibinfo{person}{Michael Felderer}, {and} \bibinfo{person}{Rudolf Ramler}.} \bibinfo{year}{2023}\natexlab{}.
\newblock \showarticletitle{The pipeline for the continuous development of artificial intelligence models—Current state of research and practice}.
\newblock \bibinfo{journal}{\emph{Journal of Systems and Software}}  \bibinfo{volume}{199} (\bibinfo{year}{2023}), \bibinfo{pages}{111615}.
\newblock


\bibitem[Testi et~al\mbox{.}(2022)]%
        {taxonomy}
\bibfield{author}{\bibinfo{person}{Matteo Testi}, \bibinfo{person}{Matteo Ballabio}, \bibinfo{person}{Emanuele Frontoni}, \bibinfo{person}{Giulio Iannello}, \bibinfo{person}{Sara Moccia}, \bibinfo{person}{Paolo Soda}, {and} \bibinfo{person}{Gennaro Vessio}.} \bibinfo{year}{2022}\natexlab{}.
\newblock \showarticletitle{MLOps: A Taxonomy and a Methodology}.
\newblock \bibinfo{journal}{\emph{IEEE Access}}  \bibinfo{volume}{10} (\bibinfo{year}{2022}), \bibinfo{pages}{63606--63618}.
\newblock
\urldef\tempurl%
\url{https://doi.org/10.1109/ACCESS.2022.3181730}
\showDOI{\tempurl}


\bibitem[Visengeriyeva et~al\mbox{.}(2023)]%
        {MLOpsPrinciples}
\bibfield{author}{\bibinfo{person}{Larysa Visengeriyeva}, \bibinfo{person}{Anja Kammer}, \bibinfo{person}{Isabel Bär}, \bibinfo{person}{Alexander Kniesz}, {and} \bibinfo{person}{Michael Plöd}.} \bibinfo{year}{2023}\natexlab{}.
\newblock \bibinfo{title}{MLOps Principles}.
\newblock \bibinfo{howpublished}{\url{https://ml-ops.org/content/mlops-principles}}.
\newblock


\bibitem[Wohlin et~al\mbox{.}(2012)]%
        {Wohlin2012}
\bibfield{author}{\bibinfo{person}{Claes Wohlin}, \bibinfo{person}{Per Runeson}, \bibinfo{person}{Martin H{\"{o}}st}, {et~al\mbox{.}}} \bibinfo{year}{2012}\natexlab{}.
\newblock \bibinfo{booktitle}{\emph{Experimentation in Software Engineering}}.
\newblock \bibinfo{publisher}{Springer}.
\newblock


\end{thebibliography}

\end{document}